\documentclass[aps,prl,superscriptaddress,nofootinbib,showkeys,twocolumn,preprintnumbers]{revtex4-1}
\pdfoutput=1
\usepackage[utf8]{inputenc}
\usepackage{uniinput}
\usepackage{graphicx}
\usepackage{color}
\usepackage{amsmath,amssymb,amsthm,mathtools,dsfont,amsfonts}
\usepackage{comment,braket,array,bold-extra,cancel,soul}
\usepackage{enumerate,xargs}
\usepackage{hyperref}
\usepackage[dvipsnames]{xcolor}
\usepackage{epstopdf}

\hypersetup{
	colorlinks = true,
	linkcolor = blue,
	anchorcolor = blue,
	citecolor = blue,
	filecolor = blue,
	urlcolor = blue
}

\newcommand{\be}{\begin{equation}}
\newcommand{\ee}{\end{equation}}
\DeclareMathOperator{\arccosh}{arccosh}
\def\({\left (}
\def\){\right )}

\newcommand{\figref}[1]{Fig.\,\ref{#1}}

\newcommand{\lp}{\left(}
\newcommand{\rp}{\right)}
\newcommand{\sig}{\sigma}
\newcommand{\vol}{\textrm{vol}}

\def\Label#1{\label{#1}%
  \smash{\hbox to0pt{\raise1ex\hbox{\tiny[#1]}\hss}}}
\def\noLabels{\let\Label=\label}
\def\nobbibitem{\let\bbibitem=\bibitem}

\begin{document}
\noLabels
\nobbibitem

\title{\Large{Emergent de Sitter Cosmology from Decaying AdS}}
\author{Souvik \surname{Banerjee}}
\author{Ulf \surname{Danielsson}}
\author{Giuseppe \surname{Dibitetto}}
\author{Suvendu \surname{Giri}}
\author{Marjorie \surname{Schillo}}
\affiliation{\it Institutionen för fysik och astronomi, Uppsala University,\\
	Box 803, SE-751 08 Uppsala, Sweden }

\begin{abstract}
\noindent
Recent developments in string compactifications demonstrate  obstructions to the simplest constructions of low energy cosmologies with positive vacuum energy. 
The existence of obstacles to creating scale-separated de Sitter solutions indicates a UV/IR puzzle for embedding cosmological vacua in a unitary theory of quantum gravity.  Motivated by this puzzle, we propose an embedding of positive energy Friedmann-Lema\^itre-Robertson-Walker (FLRW) cosmology within 
string theory.
Our proposal involves confining 4D gravity on a brane which mediates the decay from a non-supersymmetric false $\textrm{AdS}_{5}$ vacuum to a true vacuum.  In this way, it is natural for a 4D observer to experience an effective positive cosmological constant coupled to matter and radiation, avoiding the need for scale separation or a fundamental de Sitter vacuum.
\end{abstract}

\preprint{UUITP-27-18}

\maketitle


\section{Introduction}
\noindent Since the discovery of dark energy,
string theory has been faced with the challenge of reproducing a small positive vacuum energy. The dominant approach has been the reliance on a landscape of different vacua \cite{Bousso:2000xa} equipped with a transition mechanism such that the anthropic principle selects our vacuum \cite{Weinberg:1988cp}. 
This approach became calculable in string theory with the construction of KKLT \cite{Kachru:2003aw}, in which one can achieve a landscape of scale-separated vacua, with both positive and negative cosmological constant (CC), by tuning flux numbers. However, there has been a  recent debate regarding the correct application of supersymmetry-breaking and non-perturbative effects necessary in the construction of the landscape \cite{Moritz:2017xto, Sethi:2017phn, Kallosh:2018wme, Akrami:2018ylq, Kachru:2018aqn, Danielsson:2018ztv,Obied:2018sgi,Agrawal:2018own,Andriot:2018wzk}.  While the issues are not yet settled, some authors suggest that not a single rigorous string vacuum has been constructed, and further speculate that string theory abhors de Sitter space and any solution with positive vacuum energy will suffer from instabilities.

Given this lively debate,  it is reasonable to consider the possibility that neither metastability nor scale separation are achieved in string theory in the way envisioned. 
It seems, therefore, that something completely different may be needed.  In order to construct  an alternative, we will take motivation from work that received considerable attention around the turn of the millennium just before the idea of the string landscape started to flourish: braneworlds. In this context, the cosmology we see as 4D observers is not due to vacuum energy, but rather arises as an effective description on a dynamical object embedded in a higher dimensional space.

In the scenario developed by Randall and Sundrum (RS)\cite{Randall:1999ee, Randall:1999vf}, two identical $\textrm{AdS}_5$ vacua are glued together across a three-brane. 
The 5D graviton has a zero mode confined on the brane that gives rise to an effective 4D gravity despite the existence of large extra dimensions; this solves the issue of finding scale-separated vacua.  We will consider a variation of this scenario that starts with a metastable false $\textrm{AdS}_5$ vacuum that non-perturbatively decays to a supersymmetric true $\textrm{AdS}_5$ vacuum through bubble nucleation.  Here, a spherical brane separates the two phases with an inside and an outside,\footnote{Inside(outside) refers to the volume of a radial slice increasing(decreasing) as you move towards the brane.} and 4D observers confined to the brane experience an effective $\textrm{dS}_4$.  This scenario is further motivated by recent arguments that all non-supersymmetric AdS vacua must possess such an instability in a consistent theory of quantum gravity \cite{Ooguri:2016pdq, Freivogel:2016qwc}. 

\section{A world on a shell}
\begin{figure}
	\centering
	\includegraphics[width=\columnwidth]{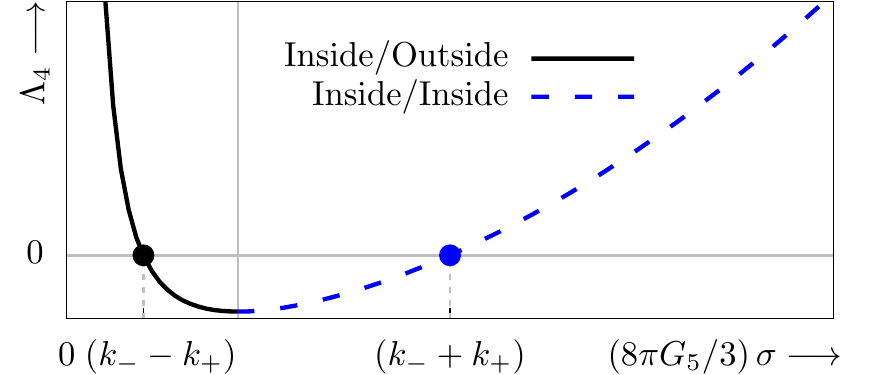}
	\caption{The $4$D CC as a function of tension.}
	\label{fig:siglam}
\end{figure}

\noindent
Let us consider the cosmology on top of an expanding bubble of true AdS₅ vacuum (with CC: $Λ_-=-6k_-²=-6/L_-²$) that has nucleated in the background of a false AdS₅ vacuum ($Λ₊=-6k₊²=-6/L_+²$), where $k_->k₊$.
While such constructions have been studied (see \emph{e.g.}\,\cite{Collins:2000yb,Bowcock:2000cq,Deruelle:2000ge}), the majority of the literature focuses on RS-like scenarios which connect two insides, or the spacetime has an exact $\mathbb{Z}₂$ symmetry.
The bubble nucleation process demands an inside/outside construction where there can be no $\mathbb{Z}₂$ symmetry across the bubble.
In global coordinates, the AdS₅ metric inside ($-$) and outside ($+$) of the bubble is given by
$ds²=-f_±(r)dt²+f_±(r)^{-1}dr²+r²dΩ²₃$,
where $f_±(r)\coloneqq1+k²_±r²$.
In terms of proper time $τ$ on the shell located at $r=a(τ)$, the induced metric takes the (FLRW) form
$ds²_{\mathrm{shell}}=-dτ²+a(τ)²dΩ²₃$.
In order for this composite spacetime to be a solution of the Einstein equations, the stress-energy tensor on the shell needs to source a jump in extrinsic curvature.  This results in a constant tension of the shell  à la Israel-Lanczos-Sen\footnote{Here we have suppressed world-volume fields that are constrained to the brane, which would appear on the RHS.  We will comment on their contribution below.} \cite{Israel:1966rt,Lanczos:1924,Sen:1924}:
\begin{equation}\label{eq:junct}
\sigma=\frac{3}{8πG_5}\left(\sqrt{k^2_{-}+\frac{1+\dot{a}^2}{a^2}}-\sqrt{k_{+}^2+\frac{1+\dot{a}^2}{a^2}}\right)\,,
\end{equation}
where the dot denotes derivative with respect to $τ$. In \figref{fig:siglam} we plot the induced CC as a function of the tension.
In general, the resulting Friedmann equation will be non-linear in the shell tension. However, when the 5D CC's, $k_±$, are large compared to the 4D Hubble parameter, the tension of the shell approaches (from below) the extremal tension which results in a flat shell
\begin{equation}\label{eq:lam4sigext}
σ_{\mathrm{ext}}=3\left(k_{-}-k₊\right)/\left(8πG_5\right)\,.
\end{equation}
Expanding \eqref{eq:junct} in $ϵ=1-σ/σ_{\mathrm{ext}}$ gives the usual Friedmann equation plus small corrections:
\begin{equation}\label{eq:friedmann_vacuum}
\frac{\dot{a}²}{a²}=-\frac{1}{a²}+\frac{8πG₄}{3}Λ₄+{\cal O}(ϵ²)\,,
\end{equation}
with the identifications
\begin{equation}\label{eq:G4Lambda4}
G₄=2k_-k_+G₅/\left(k_{-}-k_+\right)\quad\text{and}\quad Λ₄=σ_{\mathrm{ext}}-σ\,.
\end{equation}
This shows that in order to have an expanding (dS) bubble, the tension needs to be sub-extremal \emph{i.e.}\,$σ<σ_{\mathrm{ext}}$. The bubble nucleates with $\dot{a}=0$ with its radius set by the 4D CC. As a consequence, the universe starts out with a size comparable to the horizon scale, with the subsequent expansion further reducing the curvature.

The presence of mass in the bulk modifies the AdS₅ metric via $f(r)=1+k²_±r²-4G₅m_±/(πr²)$, where $m_{-(+)}$ is the mass measured inside(outside) bubble.
Through the space-space component of  the junction condition, one sees that matter in the bulk contributes a radiation term to the Friedmann equation
\begin{equation}\label{eq:Friedmann}
\frac{\dot{a}²}{a²}=-\frac{1}{a²}+\frac{8π}{3}G₄\left(Λ₄+\frac{3}{4π²}\left(\frac{m₊}{k₊}-\frac{m_-}{k_-}\right)\frac{1}{a⁴}\right)\,,
\end{equation}
where we drop terms higher order in $\epsilon$ and $m_\pm/k_\pm$. The time-time component of the junction condition (\emph{i.e.} the second Friedmann equation) can be combined with the first Friedmann equation to reproduce the 4D continuity equation on the shell. From this perspective we can identify the gravitational backreaction of bulk matter as the source of an effective energy density with a radiation equation of state on the shell.

Adding matter is a bit more subtle. 
As we have just seen, 5D matter confined to the shell yields a $1/a^4$ contribution to the Friedmann equation.
The way to get a matter contribution that dilutes with volume, $1/a^3$, is to construct massive particles as strings ending on the shell. 
The metric {\footnote{Black hole geometries sourced by uniform distribution of strings have been discussed in literature. See \cite{Herscovich:2010vr} for such a construction in flat spacetime and \cite{Chakrabortty:2011sp} for an analogous construction in AdS in a different context.}} is then given by $m_-(r)=0$ for $r<a$ and $m_+(r)=ηr$ for $r>a$,
where $\eta$  is the effective tension of the strings. This gives an effective 4D-matter with density $ρ=η/{a³k₊}$ on the shell.
As the shell climbs up the throat it eats the strings and the massive particles represented by the end points are supplied with the required potential energy to keep a constant rest mass. 

If all of these massive particles annihilate into massless radiation on the shell, by 5D energy conservation, $m_+$ just outside of the shell would be equal to the total mass of the strings that vanished. In addition $m_-$, evaluated on the shell, will increase dramatically to represent the mass that was captured by the shell. The 4D observer only feels the difference $\left(m_+-m_-\right)$, which will be determined through 4D energy conservation. In this way, all processes on the shellworld will be like shadows of processes taking place in 5D involving much larger energies.

One might wonder about the stability of the universe on top of the shell. An obvious decay channel would be another bubble of true vacuum nucleating on top of the shell as was considered in \cite{Gregory:2001dn}. However, this decay channel seems to be absent giving a strong indication in favor of the stability of these bubbles. We intend to analyze the issue of stability in detail in an upcoming work \cite{BDDGS:2018}.

Finally, we estimate the number of degrees of freedom on the shellworld centered in AdS-Schwarzschild by considering how thermal equilibrium is established. 
For an AdS-Schwarzschild metric outside the bubble with $m=m_+$,
the effective temperature just outside the shell as measured by a distant observer, at $r\gg1/k_+$, scales as $T_+^4\sim m_+k_+^2/r^4$. In the interior of the bubble, all mass is in the form of a black hole with mass $m_-$ (not necessarily equal to $m_+$) with temperature 
$T_-^4\sim m_-k_-^2/r^4$.
For the shell to be in equilibrium we need $T_-=T_+=T$, leading to $m_-k_-^2=m_+k_+^2 $.
Since $k_->k_+$, we find that $m_-<m_+$, indicating that the black hole has lost some mass due to the presence of the shell.
Using these conditions, the energy density for radiation on the shell at $a=r$ can be estimated as
$\rho \sim \left( m_+L_{+}-m_{-}L_{-}\right)/a^4=\left(L_+^3-L_-^3\right)T^4 /a^4$,
implying that the number of degrees of freedom is proportional to $ L_+^3-L_-^3 $.\footnote{The same scaling behavior can also be derived using AdS/CFT from  the Weyl anomaly of the holographically renormalized stress-energy tensor on the shell \cite{Henningson:1998gx,Balasubramanian:1999re,deHaro:2000vlm,Skenderis:2002wp}.}

\section{Shellworld gravity }
\noindent We now turn our attention to local gravitational physics on the 4D shell. We will show that while the massless graviton propagates in 5D, it has a zero-mode in its KK reduction that mediates an effective gravitational theory at low energy.  Thus, while scattering at high momenta will probe the radial direction, low energy physics will appear 4-dimensional on the brane.

\begin{figure}
	\centering
	\includegraphics[width=\columnwidth]{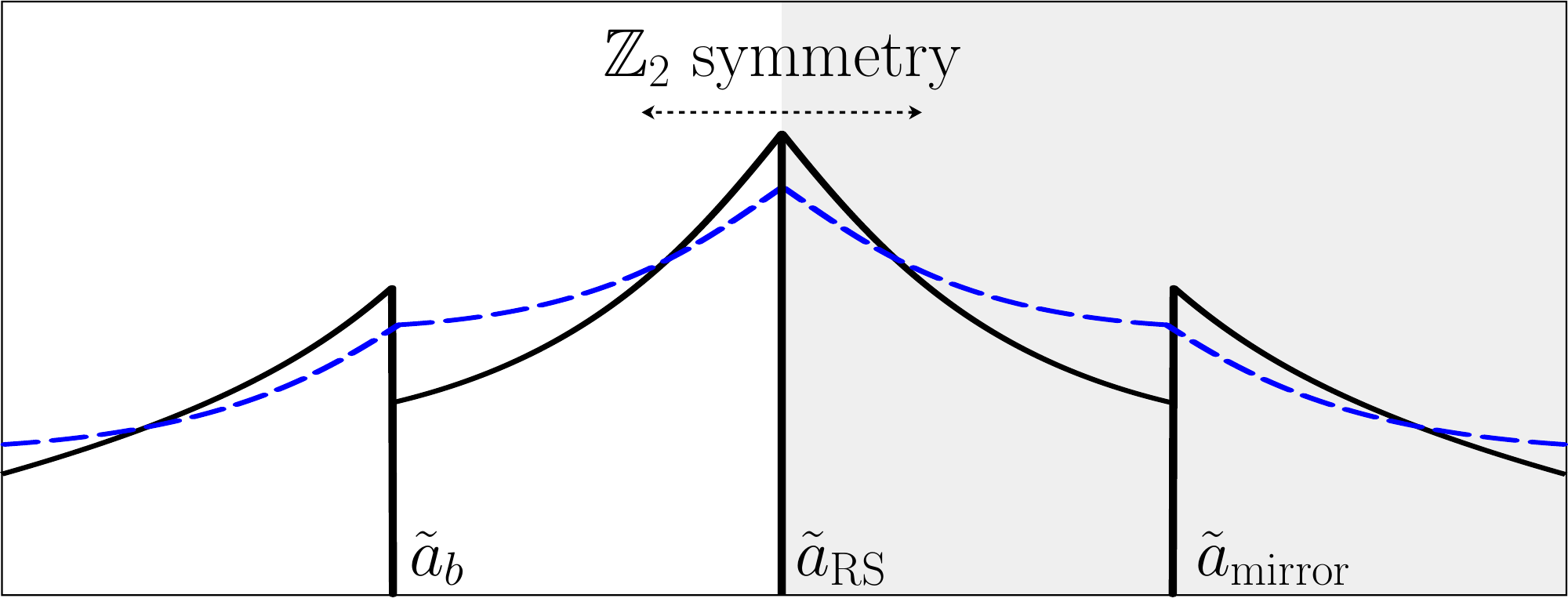}
	\caption{Potential (in Poincaré coordinates) experienced by graviton modes(solid line) and the corresponding zero mode in domain wall coordinates(dotted line).}
	\label{fig:potential}
\end{figure}

For simplicity, we consider a bubble at late times, when curvature is negligible, in the Poincar\'{e} patch, written in domain-wall coordinates:
$ds²=\tilde{a}^2(z)η_{μν}dx^μ dx^ν+dz²$,
with $\tilde{a}(z)\coloneqq\exp(zk_±)$. In RS-like constructions, $\tilde{a}(z)$ is chosen to be $\exp(zk_-)$ for $z<z_{\mathrm{RS}}$ and $\exp(-zk_+)$ for $z>z_{\mathrm{RS}}$ so that the warp factor falls off exponentially on both sides of the brane. When expressed in terms of Poincar\'{e} coordinates, this yields a volcano-shaped effective potential for the graviton modes (see the central region of \figref{fig:potential}).  

In contrast, our setup consists of an inside and an outside and therefore the warp factor increases towards the boundary. This results in a potential as shown in the unshaded region of \figref{fig:potential}. For such configurations, the zero mode of the graviton is not normalizable due to divergence at the boundary of AdS \cite{Karch:2000ct}. If we place a cut-off brane near the boundary, the non-normalizable mode causes both branes to bend \cite{Garriga:1999yh}. This effect can be interpreted as localized sources induced on the branes by the bulk modes. We identify these sources as the end points of the strings stretched between the branes. The strings result in a relation between these two sources which ensures continuity of the zero mode across the branes.
One example of this setup is depicted in \figref{fig:potential} where a RS brane at $\tilde{a}(z)=\tilde{a}_{\mathrm{RS}}$ plays the role of a cut-off brane.\footnote{In principle, the RS brane can be placed arbitrarily close to the boundary.} 
The RS brane has $\mathbb{Z}₂$ symmetry across it, which gives a mirror shell. Our shell nucleates at $\tilde{a}_b\ll\tilde{a}_{\mathrm{RS}}$ and expands with time to approach the RS brane.

Consider fluctuations to the metric of the form $g_{μν}\mapsto g_{μν}+h_{μν}$, where $h_{μν}$ is transverse and traceless. 
The radial component of $h_{μν}$ obeys a Schrödinger like equation of the form
$\left(-\partial^2_w/2+ V(w)\right)ψ(w)=(ω²-p²)ψ(w)$,
where we have changed coordinates from domain wall coordinate $z$ to the Poincaré coordinate $w$. The potential $V(w)$ is plotted in \figref{fig:potential}.

The zero mode of momentum space graviton propagators, $\chi$, with 3-momentum $p$, satisfy the equation
$\left(-p²/\tilde{a}²+\partial^2_z-4 k²\right)χ\left(p,z\right)=0$,
\cite{Padilla:2004mc} 
which has solutions 
$χ(p,z)=A(p)I₂\left[p/\left(k\tilde{a}\right)\right]+B(p)K₂\left[p/\left(k\tilde{a}\right)\right]$,
where $I₂$ and $K₂$ are Bessel functions.
Variation of the extrinsic curvature across the shell gives an additional boundary condition
$χ_-^\prime(p,\tilde{a}_b)-χ_+^\prime(p,\tilde{a}_b)-(σ/3)χ(p,\tilde{a}_b)=\Sigma_b$,
where prime denotes derivative with respect to $\tilde{a}$ and $\Sigma_b$ is a source on the shell.  
The presence of stretched strings results in a $χ(p,z)$ proportional to $K₂$ near the cut-off.
This gives
\begin{equation}
\begin{split}
\label{eq:prop1}
χ_b &= \frac{\Sigma_b\tilde{a}_b}{p}\times\left[\frac{K_1(\frac{p}{\tilde{a}_b k_-})}{K_2(\frac{p}{\tilde{a}_b k_-})}-\frac{K_1(\frac{p}{\tilde{a}_b k_+})}{K_2(\frac{p}{\tilde{a}_b k_+})}\right]^{-1}\\
&=- G₄ \frac{\Sigma_b\tilde{a}_b²}{p²+\mathcal{O}(p/a_bk_{\pm})^3},
\end{split}
\end{equation}
where $G₄ =\left(2k_+k_-\right)/\left(k_{-}-k_{+}\right)$ is the 4D Newton's constant induced on the shell by the bulk geometry. Thus, we have reproduced the correct $1/p^2$ interaction for Newtonian gravity at small $p$, including the correct constant of proportionality, $G_4$, \eqref{eq:G4Lambda4}, providing a consistency check for our scenario.
The tensor $\Sigma_b$ appearing in \eqref{eq:prop1} consists of two parts $Σ_b=Σ_b^{\rm brane}+Σ_b^{\rm str}$,
where the first term $Σ_b^{\rm brane}\coloneqq T_{μν}-Tγ_{μν}/3$ contains the contribution from the world-volume fields confined to the shell with induced metric $γ_{μν}$. This term gives a negative contribution to \eqref{eq:prop1} just as it does in the Friedmann equation.

The other piece, $Σ_{b}^{\rm str}$, arises from the bending effect of the strings on the shell.
Viewing the effect of mass as a localized deformation imparted by the endpoint of the stretched string, the contribution from strings  is of the form $\Sigma^{\text{str}}_b\sim-(\alpha_+/k_+-\alpha_-/k_-)$, where $\alpha_\pm$ is the energy carried by the strings. This follows from consistency with the Friedmann equation
when one identifies $\alpha_\pm$ with $m_\pm$. Therefore $\Sigma^{\text{str}}_b$, yields a positive contribution to $\chi_b$ in \eqref{eq:prop1}. {\it The above arguments imply that not only are the strings important for the existence of a non-vanishing zero mode, but also to ensure a well defined propagator realizing localized gravitational effects on the shellworld}.

These string sources in the bulk give an effective 5D geometry outside the bubble that mimics the CHR construction \cite{Chamblin:1999by}. Gravitational collapse of the string endpoints in 4D results in an unstable black string solution in 5D. This is the source of a longstanding problem in satisfactorily realizing black holes in a braneworld scenario \cite{Gregory:2008rf}.
However, the 5D uplift of black shells proposed in \cite{Danielsson:2017riq,Danielsson:2017pvl} avoid the instability by construction and may provide a way to realize black hole solutions on the shellworld. This is currently being investigated \cite{BDDGS:2018}.

\section{Construction in string theory}
\noindent To realize this scenario in string theory, we propose the decay of the non-supersymmetric Romans vacuum \cite{Romans:1984an} to the supersymmetric $\textrm{AdS}_5\times S^5$ vacuum via the nucleation of a spherical $(p,q)$-$5$ brane. Reducing this picture to 5D, where the $(p,q)$-$5$ brane is an $S^3$, will result in an instance of the EFT with gravitational interactions described in the previous section. The Romans vacuum is given by the reduction of type IIB over a 5D Sasaki-Einstein manifold seen as a $\textrm{U}(1)$ fibration over a $\mathbb{CP}^{3}$ base, in close analogy with the constructions in \cite{Pope:1984bd}.  The vacuum results from a non-trivial relative stretching of the fiber to the base and is supported by three-form flux.  This solution has been identified with the $\textrm{SU}(3)\times\textrm{U}(1)$ invariant critical point of 5D supergravity \cite{Gunaydin:1984qu,Gunaydin:1985cu}.

While the consistent truncation to 5D supergravity is perturbatively stable at the $\textrm{SU}(3)\times\textrm{U}(1)$ critical point \cite{Distler:1998gb,Girardello:1998pd}, it is still possible that there are tachyonic modes that have been truncated out \cite{Girardello:1999bd}.  Due to the lack of supersymmetry the instability is likely and has been taken as a foregone conclusion \cite{Pilch:2013gda}.  However, it is also possible that one may remove the offending tachyonic modes by taking an orbifold. The orbifold of the Romans vacuum will decay into $\textrm{AdS}_5\times S^5/{\mathds Z}_k$, which is in general not supersymmetric and has a bubble of nothing non-perturbative instability \cite{Horowitz:2007pr}.  However, bubbles of nothing nucleating within the true vacuum must remain inside the lightcone describing the braneworld trajectory and will therefore not affect our scenario.

In addition to the self-dual five-form flux present in the supersymmetric vacuum, the Romans vacuum has non-zero three-form flux $G_3=F_3+τH_3$, which breaks all supersymmetry and leads to the squashing of the $S^5$.  The 10D metric is given by the product $\textrm{AdS}_5\times{\cal M}_5$, where ${\cal M}_5$ is a specific fibration over $\mathbb{CP}^2$
\be
ds_{10}^2 =  \xi^2 ds_{\textrm{dS}_4}^2 + \frac{d\xi^2}{1+\xi^2/L_{\rm AdS}^2}  + L_5^2 \,\lp \nu\, ds^{2}_{\mathbb{CP}^2} + \nu^{-4}\, e_5^2 \rp \ ,
\nonumber
\ee
where $L_{\rm AdS}$ ($L_{5}$)  parametrizes the AdS (KK) scale, $\nu$ controls the relative fiber/base stretching, and $e_{5}$
is a one-form satisfying $de_{5}=J$, where $J$ is the K\"ahler form for the unit $\mathbb{CP}^2$. 

The axion $C_{(0)}$ is set to zero, while the other fluxes read
\be
\begin{split}
F_{(5)} &= \alpha_5(1+*)\vol_5\ , \\
G_{(3)} &= \alpha_3 F_{(3)} + \beta_3 H_{(3)} = d K = L_5\,e^{-3i\phi}K\wedge e_5 \ ,\\
\end{split}
\ee
where $K$ is a holomorphic two-form on $\mathbb{CP}^2$ and $\phi$ is the fiber coordinate.
Solving the 10D field equations yields
\be 
\begin{split}
{\rm SUSY}:& \quad L_{\rm AdS} = L_5\ , \quad \nu =1\ , \quad \alpha_5 = \frac{4}{g_sL_5}\ .\\
\cancel{{\rm SUSY}}:& \quad L_{\rm AdS}=\frac{2^{11/10}}{3^{3/5}}L_5\ , \quad \nu =\lp\frac23\rp^{1/5}\, , \\
&\quad \alpha_5 = \frac{108^{1/5}}{g_sL_5}\, ,\quad
 \beta_3=g_s\alpha_3 = \frac{3^{4/5}}{2^{3/10}} \frac{1}{L_5}\, .
\end{split}
\ee
For the sake of comparison we will measure the curvature in the two vacua in 5D Planck units: $M_{\rm Pl}^3=L_5^5\pi^3/(g_s^2(2\pi)^7\alpha'^4)$.  Furthermore, we are interested in decay via nucleation of five-branes which remove $G_{(3)}$ but leave $F_{(5)}$ untouched.  The flux quantization condition for $F_{(5)}$ reads
$\int_{{\cal M}_5}F_{(5)}=(4\pi^2\alpha')^2 N_5=\alpha_5L_5^5\pi^3$.
Thus, holding $N_5$ fixed and using Planck units we have:
\be
L_{\rm AdS} =\lp{N_5\over \pi}\rp^{2/3}M_{\rm Pl}^{-1}\times
\begin{cases}
1/2 & \textrm{SUSY}\,,\\
{2^{7/6}/ 3}& \cancel{\textrm{SUSY}}\,.
\end{cases}
\ee
The important result of these trivial manipulations is the hierarchy of potential energy ($V\propto L_{\rm AdS}^{-2}$) between the two vacua:
\be \label{10dhierarcy}
\text{10 dimensions:}\qquad{V_{\cancel{\rm SUSY}} \over V_{\rm SUSY}} = {9 \over 2^{13/3}}<1\ ,
\ee
indicating that the supersymmetric vacuum lies below the non-supersymmetric vacuum.

In order for a non-perturbative decay channel to exist, the $(p,q)$-$5$ brane in the theory which realizes the desired flux shift should have a tension $\sigma <\sigma_{\rm ext}$.  Using \eqref{10dhierarcy} we can compute $\sigma_{\rm ext}$ from \eqref{eq:lam4sigext}:
\be \label{sigext10d}
\sigma_{\mathrm{ext}} = M_{\rm Pl}^4\({\pi \over N_5}\)^{2/3}\(6-{9\over 2^{7/6}}\)\ .
\ee
While the precise embedding of the five-branes which mediate this decay depends on the orbifold needed to ensure perturbative stability and will be left to future work, we are able to demonstrate the existence of such a decay channel by using supergravity to obtain the tension of the requisite five-brane.  We use the superpotential and potential derived in \cite{Gunaydin:1985cu}(see also \cite{Distler:1998gb,Pilch:2000ej}):
\be \label{WandV}
\begin{split}
W(\rho,\chi) &= {1\over 4 \rho^2}\lp \cosh(2\chi)(\rho^6-2)-(3\rho^6+2)\rp \, ,\\
V(W) &= g^2\lp \frac18 \left|{\partial W \over \partial \chi}\right|^2+ \frac{1}{48} \left|\rho {\partial W \over \partial \rho}\right|^2-\frac13 \left|W\right|^2\rp\, .
\end{split}
\ee
The maximally supersymmetric critical point is at $\rho=1$, $\chi=0$, and the Romans vacuum is located at $\rho=1$, $\chi_{*}=\arccosh(2)/2$.  However one immediately notices that the hierarchy of the vacua is reversed with respect to \eqref{10dhierarcy},
i.e. $V_{\cancel{\rm SUSY}}/V_{\rm SUSY}=9/8>1$.
This is because moving in the $\chi$-direction corresponds to a deformation of the internal manifold that does not preserve $N_5$.  However, since the superpotential is linear in the fluxes, rescaling the superpotential such that the hierarchy \eqref{10dhierarcy} is recovered will also amount to holding the five-form flux fixed.  Thus, we should use the superpotential of \eqref{WandV} at the supersymmetric critical point, and $\widetilde{W}=2^{-2/3} W$ at the critical point corresponding to the Romans vacuum.

To deduce the tension of the fundamental $(p,q)5$-brane that can meditate our decay, we use the fact that there should be a BPS brane that sources the $G_{(3)}$ flux, the tension of which will be given by the junction condition for an extremal brane \eqref{eq:lam4sigext} where $k_+$ and $k_-$ are associated to $V_{+}=-g^2\widetilde{W}^2(1,\chi_*)/3$ and $V_{-}=-g^2W^2(1,0)/3$. Again, because the superpotential is linear in flux, and we have rescaled $\widetilde{W}$ such that the difference relative  to $W$ is entirely due to the change in three-form flux, using the supersymmetric values for the potential gives the effective 5D tension for the BPS brane that sources this change in flux numbers.
 
\begin{figure}
	\includegraphics[width=\columnwidth]{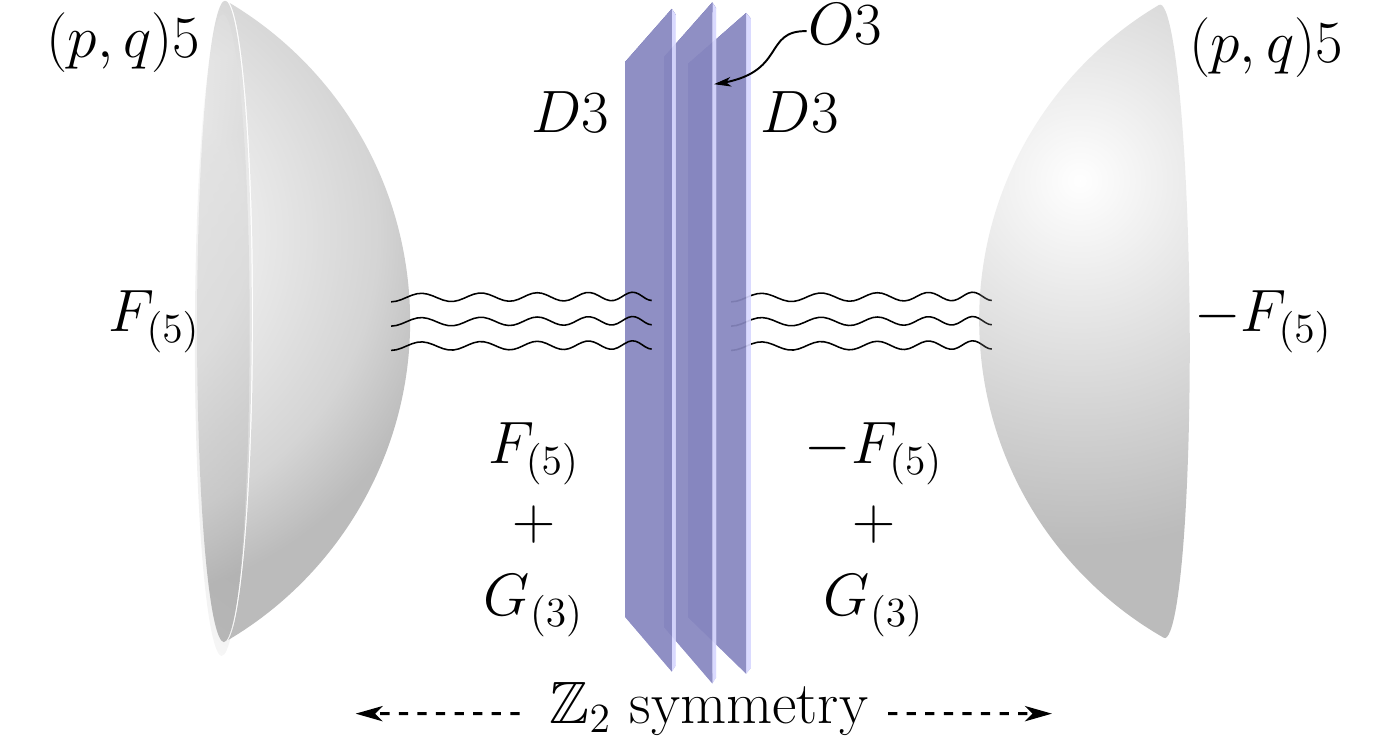}
	\caption{An artist's impression of a $(p,q)5$ braneworld.}
	\label{fig:string}
\end{figure}

Finally, in order to compare the tension of the BPS brane to the extremal tension \eqref{sigext10d}, we note that the gauge coupling is fixed to $g = 2/L_{\rm AdS}^{\rm SUSY}$ so that the potential reproduces $L_{\rm AdS}=(4\pi g_s N_5)^{1/4}\sqrt{\alpha'}$ at the supersymmetric critical point where $V_-=-3/L_{\rm AdS}^2$.  Thus, we find
\be
\begin{split}
\sig_{\rm BPS} &= 3 M_{\rm Pl}^{-3}\times {g\over 3}\(|W(1,0)|-|\widetilde{W}(1,\chi_*)|  \)\\
&= M_{\rm Pl}^4\({\pi \over N_5}\)^{2/3}\(6-{7\over 2^{2/3}}\)\, .
\end{split}
\ee
Comparing with \eqref{sigext10d} one finds $\sigma_{\rm BPS}<\sigma_{\rm ext}$ so that the fundamental brane that sources the correct charge can also facilitate the decay via a spherical bubble with finite Euclidean action.\footnote{Presence of the decay channel in this toy model does not guarantee that $\sigma$ can be tuned to near-extremal values resulting in small 4D CC and linearized Friedmann equation.}  The rescaling of $W$ that moves the brane tension away from the BPS value can also be understood as accounting for the backreaction of the moduli corresponding to the deformation of the $S^5$ in the non-supersymmetric vacuum, which is not \emph{a priori} captured by 5D supergravity.  Note that this is a non-trivial check which posed an obstacle to embedding the canonical RS scenario in string theory \cite{Kraus:1999it}.  
A stringy realization of the setup is sketched in \figref{fig:string}.
Further study of the precise brane embedding in the orbifolded geometry remains an interesting topic for future research.

\section*{Acknowledgments}
\noindent We thank A.~Retolaza, D.~Roest, L.~Tizzano, and I.~Valenzuela for stimulating discussions. 
The work of UD, GD, SG, and MS is supported by the Swedish Research Council (VR).
The work of SB is supported by the Knut and Alice Wallenberg Foundation under grant 113410212.

\bibstyle{apsrev4-1}
\bibliography{Bibliography}

\begin{thebibliography}{48}%
\makeatletter
\providecommand \@ifxundefined [1]{%
 \@ifx{#1\undefined}
}%
\providecommand \@ifnum [1]{%
 \ifnum #1\expandafter \@firstoftwo
 \else \expandafter \@secondoftwo
 \fi
}%
\providecommand \@ifx [1]{%
 \ifx #1\expandafter \@firstoftwo
 \else \expandafter \@secondoftwo
 \fi
}%
\providecommand \natexlab [1]{#1}%
\providecommand \enquote  [1]{``#1''}%
\providecommand \bibnamefont  [1]{#1}%
\providecommand \bibfnamefont [1]{#1}%
\providecommand \citenamefont [1]{#1}%
\providecommand \href@noop [0]{\@secondoftwo}%
\providecommand \href [0]{\begingroup \@sanitize@url \@href}%
\providecommand \@href[1]{\@@startlink{#1}\@@href}%
\providecommand \@@href[1]{\endgroup#1\@@endlink}%
\providecommand \@sanitize@url [0]{\catcode `\\12\catcode `\$12\catcode
  `\&12\catcode `\#12\catcode `\^12\catcode `\_12\catcode `\%12\relax}%
\providecommand \@@startlink[1]{}%
\providecommand \@@endlink[0]{}%
\providecommand \url  [0]{\begingroup\@sanitize@url \@url }%
\providecommand \@url [1]{\endgroup\@href {#1}{\urlprefix }}%
\providecommand \urlprefix  [0]{URL }%
\providecommand \Eprint [0]{\href }%
\providecommand \doibase [0]{http://dx.doi.org/}%
\providecommand \selectlanguage [0]{\@gobble}%
\providecommand \bibinfo  [0]{\@secondoftwo}%
\providecommand \bibfield  [0]{\@secondoftwo}%
\providecommand \translation [1]{[#1]}%
\providecommand \BibitemOpen [0]{}%
\providecommand \bibitemStop [0]{}%
\providecommand \bibitemNoStop [0]{.\EOS\space}%
\providecommand \EOS [0]{\spacefactor3000\relax}%
\providecommand \BibitemShut  [1]{\csname bibitem#1\endcsname}%
\let\auto@bib@innerbib\@empty
\bibitem [{\citenamefont {Bousso}\ and\ \citenamefont
  {Polchinski}(2000)}]{Bousso:2000xa}%
  \BibitemOpen
  \bibfield  {author} {\bibinfo {author} {\bibfnamefont {R.}~\bibnamefont
  {Bousso}}\ and\ \bibinfo {author} {\bibfnamefont {J.}~\bibnamefont
  {Polchinski}},\ }\href {\doibase 10.1088/1126-6708/2000/06/006} {\bibfield
  {journal} {\bibinfo  {journal} {JHEP}\ }\textbf {\bibinfo {volume} {06}},\
  \bibinfo {pages} {006} (\bibinfo {year} {2000})},\ \Eprint
  {http://arxiv.org/abs/hep-th/0004134} {arXiv:hep-th/0004134 [hep-th]}
  \BibitemShut {NoStop}%
\bibitem [{\citenamefont {Weinberg}(1989)}]{Weinberg:1988cp}%
  \BibitemOpen
  \bibfield  {author} {\bibinfo {author} {\bibfnamefont {S.}~\bibnamefont
  {Weinberg}},\ }\href {\doibase 10.1103/RevModPhys.61.1} {\bibfield  {journal}
  {\bibinfo  {journal} {Rev. Mod. Phys.}\ }\textbf {\bibinfo {volume} {61}},\
  \bibinfo {pages} {1} (\bibinfo {year} {1989})},\ \bibinfo {note}
  {[,569(1988)]}\BibitemShut {NoStop}%
\bibitem [{\citenamefont {Kachru}\ \emph {et~al.}(2003)\citenamefont {Kachru},
  \citenamefont {Kallosh}, \citenamefont {Linde},\ and\ \citenamefont
  {Trivedi}}]{Kachru:2003aw}%
  \BibitemOpen
  \bibfield  {author} {\bibinfo {author} {\bibfnamefont {S.}~\bibnamefont
  {Kachru}}, \bibinfo {author} {\bibfnamefont {R.}~\bibnamefont {Kallosh}},
  \bibinfo {author} {\bibfnamefont {A.~D.}\ \bibnamefont {Linde}}, \ and\
  \bibinfo {author} {\bibfnamefont {S.~P.}\ \bibnamefont {Trivedi}},\ }\href
  {\doibase 10.1103/PhysRevD.68.046005} {\bibfield  {journal} {\bibinfo
  {journal} {Phys. Rev.}\ }\textbf {\bibinfo {volume} {D68}},\ \bibinfo {pages}
  {046005} (\bibinfo {year} {2003})},\ \Eprint
  {http://arxiv.org/abs/hep-th/0301240} {arXiv:hep-th/0301240 [hep-th]}
  \BibitemShut {NoStop}%
\bibitem [{\citenamefont {Moritz}\ \emph {et~al.}(2018)\citenamefont {Moritz},
  \citenamefont {Retolaza},\ and\ \citenamefont {Westphal}}]{Moritz:2017xto}%
  \BibitemOpen
  \bibfield  {author} {\bibinfo {author} {\bibfnamefont {J.}~\bibnamefont
  {Moritz}}, \bibinfo {author} {\bibfnamefont {A.}~\bibnamefont {Retolaza}}, \
  and\ \bibinfo {author} {\bibfnamefont {A.}~\bibnamefont {Westphal}},\ }\href
  {\doibase 10.1103/PhysRevD.97.046010} {\bibfield  {journal} {\bibinfo
  {journal} {Phys. Rev.}\ }\textbf {\bibinfo {volume} {D97}},\ \bibinfo {pages}
  {046010} (\bibinfo {year} {2018})},\ \Eprint
  {http://arxiv.org/abs/1707.08678} {arXiv:1707.08678 [hep-th]} \BibitemShut
  {NoStop}%
\bibitem [{\citenamefont {Sethi}(2017)}]{Sethi:2017phn}%
  \BibitemOpen
  \bibfield  {author} {\bibinfo {author} {\bibfnamefont {S.}~\bibnamefont
  {Sethi}},\ }\href@noop {} {\  (\bibinfo {year} {2017})},\ \Eprint
  {http://arxiv.org/abs/1709.03554} {arXiv:1709.03554 [hep-th]} \BibitemShut
  {NoStop}%
\bibitem [{\citenamefont {Kallosh}\ \emph {et~al.}(2018)\citenamefont
  {Kallosh}, \citenamefont {Linde}, \citenamefont {McDonough},\ and\
  \citenamefont {Scalisi}}]{Kallosh:2018wme}%
  \BibitemOpen
  \bibfield  {author} {\bibinfo {author} {\bibfnamefont {R.}~\bibnamefont
  {Kallosh}}, \bibinfo {author} {\bibfnamefont {A.}~\bibnamefont {Linde}},
  \bibinfo {author} {\bibfnamefont {E.}~\bibnamefont {McDonough}}, \ and\
  \bibinfo {author} {\bibfnamefont {M.}~\bibnamefont {Scalisi}},\ }\href@noop
  {} {\  (\bibinfo {year} {2018})},\ \Eprint {http://arxiv.org/abs/1808.09428}
  {arXiv:1808.09428 [hep-th]} \BibitemShut {NoStop}%
\bibitem [{\citenamefont {Akrami}\ \emph {et~al.}(2018)\citenamefont {Akrami},
  \citenamefont {Kallosh}, \citenamefont {Linde},\ and\ \citenamefont
  {Vardanyan}}]{Akrami:2018ylq}%
  \BibitemOpen
  \bibfield  {author} {\bibinfo {author} {\bibfnamefont {Y.}~\bibnamefont
  {Akrami}}, \bibinfo {author} {\bibfnamefont {R.}~\bibnamefont {Kallosh}},
  \bibinfo {author} {\bibfnamefont {A.}~\bibnamefont {Linde}}, \ and\ \bibinfo
  {author} {\bibfnamefont {V.}~\bibnamefont {Vardanyan}},\ }\href@noop {} {\
  (\bibinfo {year} {2018})},\ \Eprint {http://arxiv.org/abs/1808.09440}
  {arXiv:1808.09440 [hep-th]} \BibitemShut {NoStop}%
\bibitem [{\citenamefont {Kachru}\ and\ \citenamefont
  {Trivedi}(2018)}]{Kachru:2018aqn}%
  \BibitemOpen
  \bibfield  {author} {\bibinfo {author} {\bibfnamefont {S.}~\bibnamefont
  {Kachru}}\ and\ \bibinfo {author} {\bibfnamefont {S.~P.}\ \bibnamefont
  {Trivedi}},\ }\href@noop {} {\  (\bibinfo {year} {2018})},\ \Eprint
  {http://arxiv.org/abs/1808.08971} {arXiv:1808.08971 [hep-th]} \BibitemShut
  {NoStop}%
\bibitem [{\citenamefont {Danielsson}\ and\ \citenamefont
  {Van~Riet}(2018)}]{Danielsson:2018ztv}%
  \BibitemOpen
  \bibfield  {author} {\bibinfo {author} {\bibfnamefont {U.~H.}\ \bibnamefont
  {Danielsson}}\ and\ \bibinfo {author} {\bibfnamefont {T.}~\bibnamefont
  {Van~Riet}},\ }\href@noop {} {\  (\bibinfo {year} {2018})},\ \Eprint
  {http://arxiv.org/abs/1804.01120} {arXiv:1804.01120 [hep-th]} \BibitemShut
  {NoStop}%
\bibitem [{\citenamefont {Obied}\ \emph {et~al.}(2018)\citenamefont {Obied},
  \citenamefont {Ooguri}, \citenamefont {Spodyneiko},\ and\ \citenamefont
  {Vafa}}]{Obied:2018sgi}%
  \BibitemOpen
  \bibfield  {author} {\bibinfo {author} {\bibfnamefont {G.}~\bibnamefont
  {Obied}}, \bibinfo {author} {\bibfnamefont {H.}~\bibnamefont {Ooguri}},
  \bibinfo {author} {\bibfnamefont {L.}~\bibnamefont {Spodyneiko}}, \ and\
  \bibinfo {author} {\bibfnamefont {C.}~\bibnamefont {Vafa}},\ }\href@noop {}
  {\  (\bibinfo {year} {2018})},\ \Eprint {http://arxiv.org/abs/1806.08362}
  {arXiv:1806.08362 [hep-th]} \BibitemShut {NoStop}%
\bibitem [{\citenamefont {Agrawal}\ \emph {et~al.}(2018)\citenamefont
  {Agrawal}, \citenamefont {Obied}, \citenamefont {Steinhardt},\ and\
  \citenamefont {Vafa}}]{Agrawal:2018own}%
  \BibitemOpen
  \bibfield  {author} {\bibinfo {author} {\bibfnamefont {P.}~\bibnamefont
  {Agrawal}}, \bibinfo {author} {\bibfnamefont {G.}~\bibnamefont {Obied}},
  \bibinfo {author} {\bibfnamefont {P.~J.}\ \bibnamefont {Steinhardt}}, \ and\
  \bibinfo {author} {\bibfnamefont {C.}~\bibnamefont {Vafa}},\ }\href@noop {}
  {\  (\bibinfo {year} {2018})},\ \Eprint {http://arxiv.org/abs/1806.09718}
  {arXiv:1806.09718 [hep-th]} \BibitemShut {NoStop}%
\bibitem [{\citenamefont {Andriot}(2018)}]{Andriot:2018wzk}%
  \BibitemOpen
  \bibfield  {author} {\bibinfo {author} {\bibfnamefont {D.}~\bibnamefont
  {Andriot}},\ }\href@noop {} {\  (\bibinfo {year} {2018})},\ \Eprint
  {http://arxiv.org/abs/1806.10999} {arXiv:1806.10999 [hep-th]} \BibitemShut
  {NoStop}%
\bibitem [{\citenamefont {Randall}\ and\ \citenamefont
  {Sundrum}(1999{\natexlab{a}})}]{Randall:1999ee}%
  \BibitemOpen
  \bibfield  {author} {\bibinfo {author} {\bibfnamefont {L.}~\bibnamefont
  {Randall}}\ and\ \bibinfo {author} {\bibfnamefont {R.}~\bibnamefont
  {Sundrum}},\ }\href {\doibase 10.1103/PhysRevLett.83.3370} {\bibfield
  {journal} {\bibinfo  {journal} {Phys. Rev. Lett.}\ }\textbf {\bibinfo
  {volume} {83}},\ \bibinfo {pages} {3370} (\bibinfo {year}
  {1999}{\natexlab{a}})},\ \Eprint {http://arxiv.org/abs/hep-ph/9905221}
  {arXiv:hep-ph/9905221 [hep-ph]} \BibitemShut {NoStop}%
\bibitem [{\citenamefont {Randall}\ and\ \citenamefont
  {Sundrum}(1999{\natexlab{b}})}]{Randall:1999vf}%
  \BibitemOpen
  \bibfield  {author} {\bibinfo {author} {\bibfnamefont {L.}~\bibnamefont
  {Randall}}\ and\ \bibinfo {author} {\bibfnamefont {R.}~\bibnamefont
  {Sundrum}},\ }\href {\doibase 10.1103/PhysRevLett.83.4690} {\bibfield
  {journal} {\bibinfo  {journal} {Phys. Rev. Lett.}\ }\textbf {\bibinfo
  {volume} {83}},\ \bibinfo {pages} {4690} (\bibinfo {year}
  {1999}{\natexlab{b}})},\ \Eprint {http://arxiv.org/abs/hep-th/9906064}
  {arXiv:hep-th/9906064 [hep-th]} \BibitemShut {NoStop}%
\bibitem [{\citenamefont {Ooguri}\ and\ \citenamefont
  {Vafa}(2017)}]{Ooguri:2016pdq}%
  \BibitemOpen
  \bibfield  {author} {\bibinfo {author} {\bibfnamefont {H.}~\bibnamefont
  {Ooguri}}\ and\ \bibinfo {author} {\bibfnamefont {C.}~\bibnamefont {Vafa}},\
  }\href {\doibase 10.4310/ATMP.2017.v21.n7.a8} {\bibfield  {journal} {\bibinfo
   {journal} {Adv. Theor. Math. Phys.}\ }\textbf {\bibinfo {volume} {21}},\
  \bibinfo {pages} {1787} (\bibinfo {year} {2017})},\ \Eprint
  {http://arxiv.org/abs/1610.01533} {arXiv:1610.01533 [hep-th]} \BibitemShut
  {NoStop}%
\bibitem [{\citenamefont {Freivogel}\ and\ \citenamefont
  {Kleban}(2016)}]{Freivogel:2016qwc}%
  \BibitemOpen
  \bibfield  {author} {\bibinfo {author} {\bibfnamefont {B.}~\bibnamefont
  {Freivogel}}\ and\ \bibinfo {author} {\bibfnamefont {M.}~\bibnamefont
  {Kleban}},\ }\href@noop {} {\  (\bibinfo {year} {2016})},\ \Eprint
  {http://arxiv.org/abs/1610.04564} {arXiv:1610.04564 [hep-th]} \BibitemShut
  {NoStop}%
\bibitem [{\citenamefont {Collins}\ and\ \citenamefont
  {Holdom}(2000)}]{Collins:2000yb}%
  \BibitemOpen
  \bibfield  {author} {\bibinfo {author} {\bibfnamefont {H.}~\bibnamefont
  {Collins}}\ and\ \bibinfo {author} {\bibfnamefont {B.}~\bibnamefont
  {Holdom}},\ }\href {\doibase 10.1103/PhysRevD.62.105009} {\bibfield
  {journal} {\bibinfo  {journal} {Phys. Rev.}\ }\textbf {\bibinfo {volume}
  {D62}},\ \bibinfo {pages} {105009} (\bibinfo {year} {2000})},\ \Eprint
  {http://arxiv.org/abs/hep-ph/0003173} {arXiv:hep-ph/0003173 [hep-ph]}
  \BibitemShut {NoStop}%
\bibitem [{\citenamefont {Bowcock}\ \emph {et~al.}(2000)\citenamefont
  {Bowcock}, \citenamefont {Charmousis},\ and\ \citenamefont
  {Gregory}}]{Bowcock:2000cq}%
  \BibitemOpen
  \bibfield  {author} {\bibinfo {author} {\bibfnamefont {P.}~\bibnamefont
  {Bowcock}}, \bibinfo {author} {\bibfnamefont {C.}~\bibnamefont {Charmousis}},
  \ and\ \bibinfo {author} {\bibfnamefont {R.}~\bibnamefont {Gregory}},\ }\href
  {\doibase 10.1088/0264-9381/17/22/313} {\bibfield  {journal} {\bibinfo
  {journal} {Class. Quant. Grav.}\ }\textbf {\bibinfo {volume} {17}},\ \bibinfo
  {pages} {4745} (\bibinfo {year} {2000})},\ \Eprint
  {http://arxiv.org/abs/hep-th/0007177} {arXiv:hep-th/0007177 [hep-th]}
  \BibitemShut {NoStop}%
\bibitem [{\citenamefont {Deruelle}\ and\ \citenamefont
  {Dolezel}(2000)}]{Deruelle:2000ge}%
  \BibitemOpen
  \bibfield  {author} {\bibinfo {author} {\bibfnamefont {N.}~\bibnamefont
  {Deruelle}}\ and\ \bibinfo {author} {\bibfnamefont {T.}~\bibnamefont
  {Dolezel}},\ }\href {\doibase 10.1103/PhysRevD.62.103502} {\bibfield
  {journal} {\bibinfo  {journal} {Phys. Rev.}\ }\textbf {\bibinfo {volume}
  {D62}},\ \bibinfo {pages} {103502} (\bibinfo {year} {2000})},\ \Eprint
  {http://arxiv.org/abs/gr-qc/0004021} {arXiv:gr-qc/0004021 [gr-qc]}
  \BibitemShut {NoStop}%
\bibitem [{\citenamefont {Israel}(1966)}]{Israel:1966rt}%
  \BibitemOpen
  \bibfield  {author} {\bibinfo {author} {\bibfnamefont {W.}~\bibnamefont
  {Israel}},\ }\href {\doibase 10.1007/BF02710419, 10.1007/BF02712210}
  {\bibfield  {journal} {\bibinfo  {journal} {Nuovo Cim.}\ }\textbf {\bibinfo
  {volume} {B44S10}},\ \bibinfo {pages} {1} (\bibinfo {year} {1966})},\
  \bibinfo {note} {[Nuovo Cim.B44,1(1966)]}\BibitemShut {NoStop}%
\bibitem [{\citenamefont {Lanczos}(1924)}]{Lanczos:1924}%
  \BibitemOpen
  \bibfield  {author} {\bibinfo {author} {\bibfnamefont {K.}~\bibnamefont
  {Lanczos}},\ }\href {\doibase 10.1002/andp.19243791403} {\bibfield  {journal}
  {\bibinfo  {journal} {Annalen der Physik}\ }\textbf {\bibinfo {volume}
  {379}},\ \bibinfo {pages} {518} (\bibinfo {year} {1924})}\BibitemShut
  {NoStop}%
\bibitem [{\citenamefont {v.~Laue}\ and\ \citenamefont {Sen}(1924)}]{Sen:1924}%
  \BibitemOpen
  \bibfield  {author} {\bibinfo {author} {\bibfnamefont {M.}~\bibnamefont
  {v.~Laue}}\ and\ \bibinfo {author} {\bibfnamefont {N.}~\bibnamefont {Sen}},\
  }\href {\doibase 10.1002/andp.19243791105} {\bibfield  {journal} {\bibinfo
  {journal} {Annalen der Physik}\ }\textbf {\bibinfo {volume} {379}},\ \bibinfo
  {pages} {252} (\bibinfo {year} {1924})}\BibitemShut {NoStop}%
\bibitem [{\citenamefont {Herscovich}\ and\ \citenamefont
  {Richarte}(2010)}]{Herscovich:2010vr}%
  \BibitemOpen
  \bibfield  {author} {\bibinfo {author} {\bibfnamefont {E.}~\bibnamefont
  {Herscovich}}\ and\ \bibinfo {author} {\bibfnamefont {M.~G.}\ \bibnamefont
  {Richarte}},\ }\href {\doibase 10.1016/j.physletb.2010.04.065} {\bibfield
  {journal} {\bibinfo  {journal} {Phys. Lett.}\ }\textbf {\bibinfo {volume}
  {B689}},\ \bibinfo {pages} {192} (\bibinfo {year} {2010})},\ \Eprint
  {http://arxiv.org/abs/1004.3754} {arXiv:1004.3754 [hep-th]} \BibitemShut
  {NoStop}%
\bibitem [{\citenamefont {Chakrabortty}(2011)}]{Chakrabortty:2011sp}%
  \BibitemOpen
  \bibfield  {author} {\bibinfo {author} {\bibfnamefont {S.}~\bibnamefont
  {Chakrabortty}},\ }\href {\doibase 10.1016/j.physletb.2011.09.112} {\bibfield
   {journal} {\bibinfo  {journal} {Phys. Lett.}\ }\textbf {\bibinfo {volume}
  {B705}},\ \bibinfo {pages} {244} (\bibinfo {year} {2011})},\ \Eprint
  {http://arxiv.org/abs/1108.0165} {arXiv:1108.0165 [hep-th]} \BibitemShut
  {NoStop}%
\bibitem [{\citenamefont {Gregory}\ and\ \citenamefont
  {Padilla}(2002)}]{Gregory:2001dn}%
  \BibitemOpen
  \bibfield  {author} {\bibinfo {author} {\bibfnamefont {R.}~\bibnamefont
  {Gregory}}\ and\ \bibinfo {author} {\bibfnamefont {A.}~\bibnamefont
  {Padilla}},\ }\href {\doibase 10.1088/0264-9381/19/2/308} {\bibfield
  {journal} {\bibinfo  {journal} {Class. Quant. Grav.}\ }\textbf {\bibinfo
  {volume} {19}},\ \bibinfo {pages} {279} (\bibinfo {year} {2002})},\ \Eprint
  {http://arxiv.org/abs/hep-th/0107108} {arXiv:hep-th/0107108 [hep-th]}
  \BibitemShut {NoStop}%
\bibitem [{\citenamefont {Banerjee}\ \emph {et~al.}()\citenamefont {Banerjee},
  \citenamefont {Danielsson}, \citenamefont {Dibitetto}, \citenamefont {Giri},\
  and\ \citenamefont {Schillo}}]{BDDGS:2018}%
  \BibitemOpen
  \bibfield  {author} {\bibinfo {author} {\bibfnamefont {S.}~\bibnamefont
  {Banerjee}}, \bibinfo {author} {\bibfnamefont {U.}~\bibnamefont
  {Danielsson}}, \bibinfo {author} {\bibfnamefont {G.}~\bibnamefont
  {Dibitetto}}, \bibinfo {author} {\bibfnamefont {S.}~\bibnamefont {Giri}}, \
  and\ \bibinfo {author} {\bibfnamefont {M.}~\bibnamefont {Schillo}},\
  }\bibfield  {booktitle} {\emph {\bibinfo {booktitle} {{in preparation}}},\
  }\href@noop {} {\ }\BibitemShut {NoStop}%
\bibitem [{\citenamefont {Henningson}\ and\ \citenamefont
  {Skenderis}(1998)}]{Henningson:1998gx}%
  \BibitemOpen
  \bibfield  {author} {\bibinfo {author} {\bibfnamefont {M.}~\bibnamefont
  {Henningson}}\ and\ \bibinfo {author} {\bibfnamefont {K.}~\bibnamefont
  {Skenderis}},\ }\href {\doibase 10.1088/1126-6708/1998/07/023} {\bibfield
  {journal} {\bibinfo  {journal} {JHEP}\ }\textbf {\bibinfo {volume} {07}},\
  \bibinfo {pages} {023} (\bibinfo {year} {1998})},\ \Eprint
  {http://arxiv.org/abs/hep-th/9806087} {arXiv:hep-th/9806087 [hep-th]}
  \BibitemShut {NoStop}%
\bibitem [{\citenamefont {Balasubramanian}\ and\ \citenamefont
  {Kraus}(1999)}]{Balasubramanian:1999re}%
  \BibitemOpen
  \bibfield  {author} {\bibinfo {author} {\bibfnamefont {V.}~\bibnamefont
  {Balasubramanian}}\ and\ \bibinfo {author} {\bibfnamefont {P.}~\bibnamefont
  {Kraus}},\ }\href {\doibase 10.1007/s002200050764} {\bibfield  {journal}
  {\bibinfo  {journal} {Commun. Math. Phys.}\ }\textbf {\bibinfo {volume}
  {208}},\ \bibinfo {pages} {413} (\bibinfo {year} {1999})},\ \Eprint
  {http://arxiv.org/abs/hep-th/9902121} {arXiv:hep-th/9902121 [hep-th]}
  \BibitemShut {NoStop}%
\bibitem [{\citenamefont {de~Haro}\ \emph {et~al.}(2001)\citenamefont
  {de~Haro}, \citenamefont {Solodukhin},\ and\ \citenamefont
  {Skenderis}}]{deHaro:2000vlm}%
  \BibitemOpen
  \bibfield  {author} {\bibinfo {author} {\bibfnamefont {S.}~\bibnamefont
  {de~Haro}}, \bibinfo {author} {\bibfnamefont {S.~N.}\ \bibnamefont
  {Solodukhin}}, \ and\ \bibinfo {author} {\bibfnamefont {K.}~\bibnamefont
  {Skenderis}},\ }\href {\doibase 10.1007/s002200100381} {\bibfield  {journal}
  {\bibinfo  {journal} {Commun. Math. Phys.}\ }\textbf {\bibinfo {volume}
  {217}},\ \bibinfo {pages} {595} (\bibinfo {year} {2001})},\ \Eprint
  {http://arxiv.org/abs/hep-th/0002230} {arXiv:hep-th/0002230 [hep-th]}
  \BibitemShut {NoStop}%
\bibitem [{\citenamefont {Skenderis}(2002)}]{Skenderis:2002wp}%
  \BibitemOpen
  \bibfield  {author} {\bibinfo {author} {\bibfnamefont {K.}~\bibnamefont
  {Skenderis}},\ }\bibfield  {booktitle} {\emph {\bibinfo {booktitle} {{The
  quantum structure of space-time and the geometric nature of fundamental
  interactions. Proceedings, RTN European Winter School, RTN 2002, Utrecht,
  Netherlands, January 17-22, 2002}}},\ }\href {\doibase
  10.1088/0264-9381/19/22/306} {\bibfield  {journal} {\bibinfo  {journal}
  {Class. Quant. Grav.}\ }\textbf {\bibinfo {volume} {19}},\ \bibinfo {pages}
  {5849} (\bibinfo {year} {2002})},\ \Eprint
  {http://arxiv.org/abs/hep-th/0209067} {arXiv:hep-th/0209067 [hep-th]}
  \BibitemShut {NoStop}%
\bibitem [{\citenamefont {Karch}\ and\ \citenamefont
  {Randall}(2001)}]{Karch:2000ct}%
  \BibitemOpen
  \bibfield  {author} {\bibinfo {author} {\bibfnamefont {A.}~\bibnamefont
  {Karch}}\ and\ \bibinfo {author} {\bibfnamefont {L.}~\bibnamefont
  {Randall}},\ }\bibfield  {booktitle} {\emph {\bibinfo {booktitle}
  {{Superstrings. Proceedings, International Conference, Strings 2000, Ann
  Arbor, USA, July 10-15, 2000}}},\ }\href {\doibase
  10.1088/1126-6708/2001/05/008} {\bibfield  {journal} {\bibinfo  {journal}
  {JHEP}\ }\textbf {\bibinfo {volume} {05}},\ \bibinfo {pages} {008} (\bibinfo
  {year} {2001})},\ \bibinfo {note} {[,140(2000)]},\ \Eprint
  {http://arxiv.org/abs/hep-th/0011156} {arXiv:hep-th/0011156 [hep-th]}
  \BibitemShut {NoStop}%
\bibitem [{\citenamefont {Garriga}\ and\ \citenamefont
  {Tanaka}(2000)}]{Garriga:1999yh}%
  \BibitemOpen
  \bibfield  {author} {\bibinfo {author} {\bibfnamefont {J.}~\bibnamefont
  {Garriga}}\ and\ \bibinfo {author} {\bibfnamefont {T.}~\bibnamefont
  {Tanaka}},\ }\href {\doibase 10.1103/PhysRevLett.84.2778} {\bibfield
  {journal} {\bibinfo  {journal} {Phys. Rev. Lett.}\ }\textbf {\bibinfo
  {volume} {84}},\ \bibinfo {pages} {2778} (\bibinfo {year} {2000})},\ \Eprint
  {http://arxiv.org/abs/hep-th/9911055} {arXiv:hep-th/9911055 [hep-th]}
  \BibitemShut {NoStop}%
\bibitem [{\citenamefont {Padilla}(2005)}]{Padilla:2004mc}%
  \BibitemOpen
  \bibfield  {author} {\bibinfo {author} {\bibfnamefont {A.}~\bibnamefont
  {Padilla}},\ }\href {\doibase 10.1088/0264-9381/22/6/011} {\bibfield
  {journal} {\bibinfo  {journal} {Class. Quant. Grav.}\ }\textbf {\bibinfo
  {volume} {22}},\ \bibinfo {pages} {1087} (\bibinfo {year} {2005})},\ \Eprint
  {http://arxiv.org/abs/hep-th/0410033} {arXiv:hep-th/0410033 [hep-th]}
  \BibitemShut {NoStop}%
\bibitem [{\citenamefont {Chamblin}\ \emph {et~al.}(2000)\citenamefont
  {Chamblin}, \citenamefont {Hawking},\ and\ \citenamefont
  {Reall}}]{Chamblin:1999by}%
  \BibitemOpen
  \bibfield  {author} {\bibinfo {author} {\bibfnamefont {A.}~\bibnamefont
  {Chamblin}}, \bibinfo {author} {\bibfnamefont {S.~W.}\ \bibnamefont
  {Hawking}}, \ and\ \bibinfo {author} {\bibfnamefont {H.~S.}\ \bibnamefont
  {Reall}},\ }\href {\doibase 10.1103/PhysRevD.61.065007} {\bibfield  {journal}
  {\bibinfo  {journal} {Phys. Rev.}\ }\textbf {\bibinfo {volume} {D61}},\
  \bibinfo {pages} {065007} (\bibinfo {year} {2000})},\ \Eprint
  {http://arxiv.org/abs/hep-th/9909205} {arXiv:hep-th/9909205 [hep-th]}
  \BibitemShut {NoStop}%
\bibitem [{\citenamefont {Gregory}(2009)}]{Gregory:2008rf}%
  \BibitemOpen
  \bibfield  {author} {\bibinfo {author} {\bibfnamefont {R.}~\bibnamefont
  {Gregory}},\ }\bibfield  {booktitle} {\emph {\bibinfo {booktitle}
  {{Proceedings, 4th Aegean Summer School: Black Holes: Mytilene, Island of
  Lesvos, Greece, September 17-22, 2007}}},\ }\href {\doibase
  10.1007/978-3-540-88460-6_7} {\bibfield  {journal} {\bibinfo  {journal}
  {Lect. Notes Phys.}\ }\textbf {\bibinfo {volume} {769}},\ \bibinfo {pages}
  {259} (\bibinfo {year} {2009})},\ \Eprint {http://arxiv.org/abs/0804.2595}
  {arXiv:0804.2595 [hep-th]} \BibitemShut {NoStop}%
\bibitem [{\citenamefont {Danielsson}\ \emph {et~al.}(2017)\citenamefont
  {Danielsson}, \citenamefont {Dibitetto},\ and\ \citenamefont
  {Giri}}]{Danielsson:2017riq}%
  \BibitemOpen
  \bibfield  {author} {\bibinfo {author} {\bibfnamefont {U.~H.}\ \bibnamefont
  {Danielsson}}, \bibinfo {author} {\bibfnamefont {G.}~\bibnamefont
  {Dibitetto}}, \ and\ \bibinfo {author} {\bibfnamefont {S.}~\bibnamefont
  {Giri}},\ }\href {\doibase 10.1007/JHEP10(2017)171} {\bibfield  {journal}
  {\bibinfo  {journal} {JHEP}\ }\textbf {\bibinfo {volume} {10}},\ \bibinfo
  {pages} {171} (\bibinfo {year} {2017})},\ \Eprint
  {http://arxiv.org/abs/1705.10172} {arXiv:1705.10172 [hep-th]} \BibitemShut
  {NoStop}%
\bibitem [{\citenamefont {Danielsson}\ and\ \citenamefont
  {Giri}(2017)}]{Danielsson:2017pvl}%
  \BibitemOpen
  \bibfield  {author} {\bibinfo {author} {\bibfnamefont {U.}~\bibnamefont
  {Danielsson}}\ and\ \bibinfo {author} {\bibfnamefont {S.}~\bibnamefont
  {Giri}},\ }\href@noop {} {\  (\bibinfo {year} {2017})},\ \Eprint
  {http://arxiv.org/abs/1712.00511} {arXiv:1712.00511 [hep-th]} \BibitemShut
  {NoStop}%
\bibitem [{\citenamefont {Romans}(1985)}]{Romans:1984an}%
  \BibitemOpen
  \bibfield  {author} {\bibinfo {author} {\bibfnamefont {L.~J.}\ \bibnamefont
  {Romans}},\ }\href {\doibase 10.1016/0370-2693(85)90479-4} {\bibfield
  {journal} {\bibinfo  {journal} {Phys. Lett.}\ }\textbf {\bibinfo {volume}
  {153B}},\ \bibinfo {pages} {392} (\bibinfo {year} {1985})}\BibitemShut
  {NoStop}%
\bibitem [{\citenamefont {Pope}\ and\ \citenamefont
  {Warner}(1985)}]{Pope:1984bd}%
  \BibitemOpen
  \bibfield  {author} {\bibinfo {author} {\bibfnamefont {C.~N.}\ \bibnamefont
  {Pope}}\ and\ \bibinfo {author} {\bibfnamefont {N.~P.}\ \bibnamefont
  {Warner}},\ }\href {\doibase 10.1016/0370-2693(85)90992-X} {\bibfield
  {journal} {\bibinfo  {journal} {Phys. Lett.}\ }\textbf {\bibinfo {volume}
  {150B}},\ \bibinfo {pages} {352} (\bibinfo {year} {1985})}\BibitemShut
  {NoStop}%
\bibitem [{\citenamefont {Gunaydin}\ \emph {et~al.}(1985)\citenamefont
  {Gunaydin}, \citenamefont {Romans},\ and\ \citenamefont
  {Warner}}]{Gunaydin:1984qu}%
  \BibitemOpen
  \bibfield  {author} {\bibinfo {author} {\bibfnamefont {M.}~\bibnamefont
  {Gunaydin}}, \bibinfo {author} {\bibfnamefont {L.~J.}\ \bibnamefont
  {Romans}}, \ and\ \bibinfo {author} {\bibfnamefont {N.~P.}\ \bibnamefont
  {Warner}},\ }\href {\doibase 10.1016/0370-2693(85)90361-2} {\bibfield
  {journal} {\bibinfo  {journal} {Phys. Lett.}\ }\textbf {\bibinfo {volume}
  {154B}},\ \bibinfo {pages} {268} (\bibinfo {year} {1985})}\BibitemShut
  {NoStop}%
\bibitem [{\citenamefont {Gunaydin}\ \emph {et~al.}(1986)\citenamefont
  {Gunaydin}, \citenamefont {Romans},\ and\ \citenamefont
  {Warner}}]{Gunaydin:1985cu}%
  \BibitemOpen
  \bibfield  {author} {\bibinfo {author} {\bibfnamefont {M.}~\bibnamefont
  {Gunaydin}}, \bibinfo {author} {\bibfnamefont {L.~J.}\ \bibnamefont
  {Romans}}, \ and\ \bibinfo {author} {\bibfnamefont {N.~P.}\ \bibnamefont
  {Warner}},\ }\href {\doibase 10.1016/0550-3213(86)90237-3} {\bibfield
  {journal} {\bibinfo  {journal} {Nucl. Phys.}\ }\textbf {\bibinfo {volume}
  {B272}},\ \bibinfo {pages} {598} (\bibinfo {year} {1986})}\BibitemShut
  {NoStop}%
\bibitem [{\citenamefont {Distler}\ and\ \citenamefont
  {Zamora}(1999)}]{Distler:1998gb}%
  \BibitemOpen
  \bibfield  {author} {\bibinfo {author} {\bibfnamefont {J.}~\bibnamefont
  {Distler}}\ and\ \bibinfo {author} {\bibfnamefont {F.}~\bibnamefont
  {Zamora}},\ }\href {\doibase 10.4310/ATMP.1998.v2.n6.a6} {\bibfield
  {journal} {\bibinfo  {journal} {Adv. Theor. Math. Phys.}\ }\textbf {\bibinfo
  {volume} {2}},\ \bibinfo {pages} {1405} (\bibinfo {year} {1999})},\ \Eprint
  {http://arxiv.org/abs/hep-th/9810206} {arXiv:hep-th/9810206 [hep-th]}
  \BibitemShut {NoStop}%
\bibitem [{\citenamefont {Girardello}\ \emph {et~al.}(1998)\citenamefont
  {Girardello}, \citenamefont {Petrini}, \citenamefont {Porrati},\ and\
  \citenamefont {Zaffaroni}}]{Girardello:1998pd}%
  \BibitemOpen
  \bibfield  {author} {\bibinfo {author} {\bibfnamefont {L.}~\bibnamefont
  {Girardello}}, \bibinfo {author} {\bibfnamefont {M.}~\bibnamefont {Petrini}},
  \bibinfo {author} {\bibfnamefont {M.}~\bibnamefont {Porrati}}, \ and\
  \bibinfo {author} {\bibfnamefont {A.}~\bibnamefont {Zaffaroni}},\ }\href
  {\doibase 10.1088/1126-6708/1998/12/022} {\bibfield  {journal} {\bibinfo
  {journal} {JHEP}\ }\textbf {\bibinfo {volume} {12}},\ \bibinfo {pages} {022}
  (\bibinfo {year} {1998})},\ \Eprint {http://arxiv.org/abs/hep-th/9810126}
  {arXiv:hep-th/9810126 [hep-th]} \BibitemShut {NoStop}%
\bibitem [{\citenamefont {Girardello}\ \emph {et~al.}(2000)\citenamefont
  {Girardello}, \citenamefont {Petrini}, \citenamefont {Porrati},\ and\
  \citenamefont {Zaffaroni}}]{Girardello:1999bd}%
  \BibitemOpen
  \bibfield  {author} {\bibinfo {author} {\bibfnamefont {L.}~\bibnamefont
  {Girardello}}, \bibinfo {author} {\bibfnamefont {M.}~\bibnamefont {Petrini}},
  \bibinfo {author} {\bibfnamefont {M.}~\bibnamefont {Porrati}}, \ and\
  \bibinfo {author} {\bibfnamefont {A.}~\bibnamefont {Zaffaroni}},\ }\href
  {\doibase 10.1016/S0550-3213(99)00764-6} {\bibfield  {journal} {\bibinfo
  {journal} {Nucl. Phys.}\ }\textbf {\bibinfo {volume} {B569}},\ \bibinfo
  {pages} {451} (\bibinfo {year} {2000})},\ \Eprint
  {http://arxiv.org/abs/hep-th/9909047} {arXiv:hep-th/9909047 [hep-th]}
  \BibitemShut {NoStop}%
\bibitem [{\citenamefont {Pilch}\ and\ \citenamefont
  {Yoo}(2013)}]{Pilch:2013gda}%
  \BibitemOpen
  \bibfield  {author} {\bibinfo {author} {\bibfnamefont {K.}~\bibnamefont
  {Pilch}}\ and\ \bibinfo {author} {\bibfnamefont {I.}~\bibnamefont {Yoo}},\
  }\href {\doibase 10.1007/JHEP09(2013)124} {\bibfield  {journal} {\bibinfo
  {journal} {JHEP}\ }\textbf {\bibinfo {volume} {09}},\ \bibinfo {pages} {124}
  (\bibinfo {year} {2013})},\ \Eprint {http://arxiv.org/abs/1305.0295}
  {arXiv:1305.0295 [hep-th]} \BibitemShut {NoStop}%
\bibitem [{\citenamefont {Horowitz}\ \emph {et~al.}(2008)\citenamefont
  {Horowitz}, \citenamefont {Orgera},\ and\ \citenamefont
  {Polchinski}}]{Horowitz:2007pr}%
  \BibitemOpen
  \bibfield  {author} {\bibinfo {author} {\bibfnamefont {G.~T.}\ \bibnamefont
  {Horowitz}}, \bibinfo {author} {\bibfnamefont {J.}~\bibnamefont {Orgera}}, \
  and\ \bibinfo {author} {\bibfnamefont {J.}~\bibnamefont {Polchinski}},\
  }\href {\doibase 10.1103/PhysRevD.77.024004} {\bibfield  {journal} {\bibinfo
  {journal} {Phys. Rev.}\ }\textbf {\bibinfo {volume} {D77}},\ \bibinfo {pages}
  {024004} (\bibinfo {year} {2008})},\ \Eprint {http://arxiv.org/abs/0709.4262}
  {arXiv:0709.4262 [hep-th]} \BibitemShut {NoStop}%
\bibitem [{\citenamefont {Pilch}\ and\ \citenamefont
  {Warner}(2000)}]{Pilch:2000ej}%
  \BibitemOpen
  \bibfield  {author} {\bibinfo {author} {\bibfnamefont {K.}~\bibnamefont
  {Pilch}}\ and\ \bibinfo {author} {\bibfnamefont {N.~P.}\ \bibnamefont
  {Warner}},\ }\href {\doibase 10.1016/S0370-2693(00)00796-6} {\bibfield
  {journal} {\bibinfo  {journal} {Phys. Lett.}\ }\textbf {\bibinfo {volume}
  {B487}},\ \bibinfo {pages} {22} (\bibinfo {year} {2000})},\ \Eprint
  {http://arxiv.org/abs/hep-th/0002192} {arXiv:hep-th/0002192 [hep-th]}
  \BibitemShut {NoStop}%
\bibitem [{\citenamefont {Kraus}(1999)}]{Kraus:1999it}%
  \BibitemOpen
  \bibfield  {author} {\bibinfo {author} {\bibfnamefont {P.}~\bibnamefont
  {Kraus}},\ }\href {\doibase 10.1088/1126-6708/1999/12/011} {\bibfield
  {journal} {\bibinfo  {journal} {JHEP}\ }\textbf {\bibinfo {volume} {12}},\
  \bibinfo {pages} {011} (\bibinfo {year} {1999})},\ \Eprint
  {http://arxiv.org/abs/hep-th/9910149} {arXiv:hep-th/9910149 [hep-th]}
  \BibitemShut {NoStop}%
\end{thebibliography}%
\end{document}